\documentclass{article}

\usepackage{spconf,amsmath,graphicx}
\usepackage{amsfonts}
\usepackage{hyperref}
\usepackage{amssymb}
\usepackage{color}

\title{Training neural audio classifiers with few data}
\name
{Jordi Pons$^{\star\,\dagger}$\thanks{Work partially funded by the Maria de Maeztu Programme (MDM-2015-0502). J. Pons and X. Serra are grateful to NVidia for the donated GPUs.} \qquad Joan Serr\`{a}$^\star$ \qquad Xavier Serra$^\dagger$}

\address{$^\star$ Telef\'{o}nica Research, Barcelona\\$^\dagger$ Music Technology Group, Universitat Pompeu Fabra, Barcelona}

\begin{document}
\ninept

\maketitle

\begin{abstract} 
We investigate supervised learning strategies that improve the training of neural network audio classifiers on small annotated collections. In particular, we study whether (i)~a naive regularization of the solution space, (ii)~prototypical networks, (iii)~transfer learning, or (iv)~their combination, can foster deep learning models to better leverage a small amount of training examples. To this end, we evaluate (i--iv)~for the tasks of acoustic event recognition and acoustic scene classification, considering from 1 to 100~labeled examples per class. Results indicate that transfer learning is a powerful strategy in such scenarios, but prototypical networks show promising results when one does not count with external or validation data. 
\end{abstract}


%%%%%%%%%%%%%%%%%%%%%%%%%%%%%%%%%%%%%%%%%%%%%%%%%%%%%%%%%%%%%%%%%%%%%%%%%%%%%%%%%%%%%%%%%%%%%

\section{Introduction}
\label{sec:intro}

It exists a prominent corpus of research assuming that sizable amounts of annotated audio data are available for training end-to-end classifiers~\cite{hershey2017cnn, salamon2017deep, pons2017timbre, pons2017end}. These studies are mostly based on publicly-available datasets, where each class typically contains more than 100 audio examples~\cite{kaggle, fonseca2017freesound, mesaros2016tut, salamon2014dataset, gemmeke2017audio}. 
Contrastingly, only few works study the problem of training neural audio classifiers with few audio examples (for instance, less than 10 per class)~\cite{bocharov2017k,tilk2018make,morfi2018deep, morfi2018data}. 
In this work, we study how a number of neural network architectures perform in such situation.
Two primary reasons motivate our work: (i) given that humans are able to learn novel concepts from few examples, we aim to quantify up to what extent such behavior is possible in current neural machine listening systems; and (ii) provided that data curation processes are tedious and expensive, it is unreasonable to assume that sizable amounts of annotated audio are always available for training neural network classifiers. 

The challenge of training neural networks with few audio data has been previously addressed. For example, Morfi and Stowell~\cite{morfi2018deep} 
approached the problem via factorising an audio transcription task into two intermediate 
sub-tasks: event and tag detection. 
Another way to approach the problem is by leveraging additional data sources, like in unsupervised and semi-supervised frameworks where non-labelled data is also utilized~\cite{jansen2017unsupervised,lee2009unsupervised,xu2017unsupervised}. 
Transfer learning is a popular way to exploit such additional data sources~\cite{kunze2017transfer,choi2017transfer}, and it has been used to construct acoustic models for low-resource languages~\cite{ghoshal2013multilingual,huang2013cross}, to adapt generative adversarial networks to new languages and noise types~\cite{Pascual18ICASSP}, or to transfer knowledge from the visual to the audio domain~\cite{aytar2016soundnet}.
An additional alternative is to use data augmentation, which has proven to be very effective for audio classification tasks~\cite{salamon2017deep,mun2017generative}.  
However, in this work, we center our efforts into exploiting additional data resources with transfer learning. This, according to our view, has three main advantages: (i)~differently to data augmentation, it allows leveraging external sources of data; (ii)~it exists a rich set of techniques for learning transferable representations~\cite{xu2017unsupervised,kunze2017transfer,choi2017transfer,huang2013cross,aytar2016soundnet}; and (iii)~transfer learning can always be further extended with data augmentation. 

In parallel to previous works, the machine learning community has been developing methods for learning novel classes from few training instances, an area known as few-shot learning~\cite{tilk2018make,snell2017prototypical,ravi2016optimization,vinyals2016matching}.
These methods aim to build a classifier that generalizes to new classes not seen during training, given only
a small number of training examples for each new class. Differently to few-shot learning, the models we study do not generalize to new classes. Instead, we assume a fixed taxonomy during both training and prediction. Still, we derive inspiration from few-shot learning for their capacity to learn from few training data.
A popular approach to few-shot learning is metric learning, which aims to learn representations that preserve the class neighborhood structure so that simple distances can be measured in a learnt space~\cite{snell2017prototypical,vinyals2016matching}.
Such methods have been mostly used for image classification, and are very appealing due to their simplicity and yet powerful performance on several benchmarks.

In our study we consider prototypical networks~\cite{snell2017prototypical}, a metric learning approach that is based on computing distances against class-based prototypes defined in a learnt embedding space (one can think of it as a nearest-neigbour classifier trained end-to-end). Our aim is to study if prototypical networks are capable to generalize better than raw deep learning models when trained on small data.
To the best of our knowledge, only Tilk~\cite{tilk2018make} has explored metric learning methods for constructing neural audio classifiers. 
He used the last layer features of a siamese network~\cite{bromley1994signature}, an alternative metric learning approach, as input to an SVM classifier.
Therefore, our work can be considered the first one to employ prototypical networks for audio. 
Besides, back in the 80's, Kohonen~\cite{kohonen1988neural} proposed a distance-based model for speech recognition called learning vector quantization (LVQ), which is closely related to prototypical networks. However, LVQ is not designed to learn from few data and, furthermore, does not exploit the powerful non-linear mapping that neural networks can provide. 

In this work, we investigate which strategies can provide a performance boost when neural network audio classifiers are trained with few data. These are evaluated under different low-data situations, which we describe in section~\ref{sec:method}. Firstly, we consider the regularization of the traditional deep learning pipeline (section~\ref{sec:sl}). Next, we consider prototypical networks, with the aim to showcase the potential of metric learning-based classifiers coming from the few-shot learning literature (section~\ref{sec:proto}). Finally, we consider transfer learning as a canonical way to leverage external sources of audio data (section~\ref{sec:transfer}). Results are presented in section~\ref{sec:results} and, to conclude, we provide further discussion in section~\ref{sec:discuss}.
We use publicly-available data sets, and share the code to reproduce our experiments at 
\href{http://github.com/jordipons/neural-classifiers-with-few-audio}{github.com/jordipons/neural-classifiers-with-few-audio}.

%%%%%%%%%%%%%%%%%%%%%%%%%%%%%%%%%%%%%%%%%%%%%%%%%%%%%%%%%%%%%%%%%%%%%%%%%%%%%%%%%%%%%%%%%%

\section{Methodology}
\label{sec:method}

\subsection{Data: Where is the validation set?}

The focus of this work is to investigate which neural network-based strategies perform best in the low-data regime. To do so, we simulate classification scenarios having only $n$ randomly selected training audios per class, $n\in\{1,2,5,10,20,50,100\}$. Since results of the same repeated experiment might vary depending on which audios are selected, we run each experiment $m$ times per fold of data, and report average accuracy scores across runs and folds. Specifically: $m=20$ when $n\in\{1, 2\}$, $m=10$ when $n\in\{5, 10\}$, and $m=5$ when $n\in\{20, 50, 100\}$.

We run the study for both the tasks of acoustic event recognition and acoustic scene classification. For acoustic event recognition, we employ the UrbanSound8K dataset (US8K)~\cite{salamon2014dataset}, featuring 8,732 urban sounds divided into 10~classes and 10~folds (with roughly 1000 instances per class). For acoustic scene classification, we resort to the TUT dataset (ASC-TUT)~\cite{mesaros2016tut,mesaros2017dcase}, featuring 4,680 audio segments for training and 1,620 for evaluation, of 10\,s each, divided into 15~classes (with 312 instances per class).
Furthermore, and we consider this a crucial aspect of our work, we assume that data is so scarce that it is unreasonable to presuppose the existence of a validation set for deciding when to stop training. As a result of such constraint, the following sections also describe the rules we use to decide when to stop training. Besides reducing the train set size and not utilizing any validation set, we keep the original partitions to compare our results with previous works. 

\subsection{Baselines}
\label{sec:baselines}

To put results into context, we employ several baselines aiming to describe lower and upper bounds for the two tasks we consider:
\begin{itemize}

\item \textbf{Random guess:} A model picking a class at random, which scores 9.99\% accuracy for US8K and 6.66\% for ASC-TUT. 

\item \textbf{Nearest-neighbor MFCCs:} A model based on a simple nearest-neighbor classifier using the cosine distance over MFCC features. The feature vector is constructed from 20~MFCCs, their $\Delta$s, and $\Delta\Delta$s. We compute their mean and standard deviation through time, with the resulting feature vector per audio clip being of size 120. 

\item \textbf{Salamon and Bello~\cite{salamon2017deep} (SB-CNN):} This model achieves state-of-the-art results for US8K. When trained with all US8K training data it achieves an average accuracy score of 73\% across folds (79\% with data augmentation). SB-CNN is an AlexNet-like model that consists of 3 convolutional layers with filters of $5\!\times\!5$, interleaved with max-pool layers. The resulting feature map is connected to a softmax output via a dense layer of 64~units. %They also use ReLUs and batch-norm. 
To assess how standard deep learning models would perform when learning their weights from scratch with small data, we train this baseline for all $n$.% (see Fig.\ref{fig:results}).

\item \textbf{Han et al.~\cite{han2017convolutional} and Mun et al.~\cite{mun2017generative}:} These models achieve state-of-the-art performance for ASC-TUT. When trained with all ASC-TUT training data, they achieve an accuracy score of 80.4\% using an ensemble~\cite{han2017convolutional}, and 83.3\% using an ensemble trained with GAN-based data augmentation~\cite{mun2017generative}.

\end{itemize}

\subsection{Training details}
\label{sec:traindetail}

Unless stated otherwise, the sections below follow the same experimental setup. Following common practice \cite{salamon2017deep,pons2018randomly}, inputs are set to be log-mel spectrogram patches of 128\,bins\,$\times$\,3\,s (128 frames)\footnote{STFT parameters: \textit{window\_size=hop\_size}=1024 and \textit{fs}=44.1\,kHz.}. For US8K, when audio clips are shorter than 3\,s, we `repeat-pad' spectrograms in order to meet the model's input size. That is, the original short signal is repeated up to create a 3\,s signal. 
During training, data are randomly sampled from the original log-mel spectrograms following the previous rules. However, during prediction, if sounds are longer than 3\,s, several predictions are computed by a moving window of 1\,s and then averaged. 
We use ReLUs and a batch size of 256. Learning proceeds via minimizing the cross-entropy loss with vanilla stochastic gradient descent (SGD) at a rate of 0.1. We stabilize learning with gradient clipping, that is, we rescale the gradients so that their L2~norm does never exceed a threshold of 5.
%Gradients are clipped when their global l2-norm exceeds a thershold of 5. %clip_by_global_norm(gradients, 5.0)

%%%%%%%%%%%%%%%%%%%%%%%%%%%%%%%%%%%%%%%%%%%%%%%%%%%%%%%%%%%%%%%%%%%%%%%%%%%%%%%%%%%%%%%%%

\section{Audio classification with few data}

\subsection{Regularized models}
\label{sec:sl}

In a first set of experiments, we showcase the limitations of the commonly used deep learning pipeline when training data are scarce. An ordinary approach to avoid overfitting in such cases is to use regularization. Following this idea, we consider two architectures. The first one, VGG, is meant to keep the model highly expressive while introducing as much regularization as possible. The second one, TIMBRE, strongly regularizes the set of possible solutions via domain-knowledge informed architectural choices.

\begin{itemize}

\item \textbf{VGG~\cite{hershey2017cnn,choi2017transfer}:} This is a computer vision architecture designed to make minimal assumptions regarding which are the local stationarities of the signal, so that any  structure  can  be  learnt  via  hierarchically combining small-context representations. To this end, it is common to utilize a deep stack of small $3\!\times\!3$ filters (in our case 5~layers, each having only 32~filters), combined with max-pool layers (in our case of $2\times\!2$). We further employ a final dense layer with a softmax activation that adapts the feature map size to the number of output classes, ELUs as non-linearities~\cite{choi2017transfer}, and batch norm.

\item \textbf{TIMBRE~\cite{pons2017timbre,pons2018randomly}:} This model is designed to learn timbral representations while keeping the model as small as possible. We use a single-layer convolutional neural network (CNN) with vertical filters of 108\,bins $\times$ 7~frames. A softmax output is computed from the maximum values present in each CNN feature map and, therefore, the model has as many filters as output classes. TIMBRE is possibly the smallest CNN one can imagine for an audio classification task, provided that it only has a single `timbral' filter per class~\cite{pons2017timbre}. % (capable to represent pitch-invariant timbral signatures).

\end{itemize}

Note that the studied VGG and TIMBRE models, of approximately 50\,k and 10\,k parameters, respectively, are much smaller than the state-of-the-art SB-CNN model, which has approximately 250\,k parameters. Besides the regularization we introduce via minimizing the model size, we employ L2-regularization, with a penalty factor of 0.001, and make use of 50\% dropout whenever a dense layer is present (only VGG and SB-CNN models have dense layers). 
Finally, and given that no validation set is available, we empirically find that (early) stopping training after 200 epochs provides good results for all studied models that are not based on prototypical networks.%standard deep learning, regularized, and \mbox{transfer learning models.}

%%%%%%%%%%%%%%%%%%%%%%%%%%%%%%%%%%%%%%%%%%%%%%%%%%%%%%%%%%%%%%%%%%%%%%%%%%%%%%%%%%%%%%%

\subsection{Prototypical networks}
\label{sec:proto}

In the next set of experiments, we study how prototypical networks can be exploited for audio classification with few examples. 
As mentioned, prototypical networks are based on learning a latent metric space in which classification can be performed by computing distances to prototype representations of each class. 
Prototypes ${\mu}_k$ are mean vectors of the embedded support data belonging to class $k$. That is, given input samples $x_i$:
\begin{equation*}
\mu_k = \frac{1}{|S_k|} \sum_{{x}_i\in S_k} f_{\phi}({x}_i) ,
\end{equation*}
where $f_{\phi}$ is parametrized by a neural network. In our case, we use a VGG as in section~\ref{sec:sl}, but with 128~filters per layer and a final linear layer instead of a softmax. The prototype vector ${\mu}_k\in \mathbb{R}^{D}$ is set to have an embedding size of $D=10$. 
In this work, the support set $S_k$ to compute each class' prototype is conformed by 5~randomly selected patches from the train set belonging to class $k$. These same sounds will be then reused to train $f_{\phi}$. 

Prototypical networks produce a distribution over classes for a query point ${x}_i$ based on a softmax over distances to the prototypes in the embedding space:
\begin{equation*}
p_k({x}_i) = \frac{e^{-d(f_\phi({x}_i),{\mu}_k)}}{\sum_{k'} e^{-d(f_\phi({x}_i),{\mu}_{k'})}} ,
\end{equation*}
where $d$ is any suitable distance measure. We here use the Euclidean distance as we found it to outperform the cosine distance for our tasks, see Appendix~\ref{app:B}.
The training of $f_{\phi}$ is based on minimizing the negative log-probability of the true class via SGD. Training epochs are formed by batches of 5 random patches per class, and we backpropagate until the train set accuracy does not improve for 200~epochs. Note that this stop criteria does not utilize a validation set, only the train set. Differently from the common supervised learning pipeline, we found overfitting not to be an issue with prototypical networks, which is an important point to consider when training models with few data. Actually, monitoring how well the model is able to separate the training data in the embedding space, through measuring train set accuracy, was an effective way to assess how discriminative such space is.
Although this stop criteria could promote overfitting the train set, in section~\ref{sec:results} we show that prototypical networks' generalization capabilities are still above the ones of the raw deep learning pipeline. Further discussion on the generalization capabilities of prototypical networks is available at Appendix~\ref{app:C}.

%%%%%%%%%%%%%%%%%%%%%%%%%%%%%%%%%%%%%%%%%%%%%%%%%%%%%%%%%%%%%%%%%%%%%%%%%%%%%%%%%%%%%%%%%

\subsection{Transfer learning}
\label{sec:transfer}

In our final set of experiments, we assess the effectiveness of canonical transfer learning strategies. For that, we use a VGG model pre-trained with Audioset~\cite{hershey2017cnn,gemmeke2017audio}, a dataset conformed by 2\,M YouTube audios that was designed for training acoustic event recognition models. As a result, note that our source and target tasks are the same for US8K (all US8K classes have a direct correspondence in Audioset), but are different for ASC-TUT (only 5 out of 15 classes resemble Audioset classes).
The pre-trained Audioset model\footnote{\href{https://github.com/tensorflow/models/tree/master/research/audioset}{github.com/tensorflow/models/tree/master/research/audioset}} is composed of 6~convolutional layers with filters of $3\!\times\!3$, interleaved with max-pooling layers of $2\!\times\!2$, followed by 3~dense layers of 4096, 4096, and 128 units, respectively.
Inputs are log-mel spectrogram patches of 64 bins $\times$ 1\,s (96 frames)\footnote{STFT parameters: \textit{window\_size}=400, \textit{hop\_size}=160 and \textit{fs}=16\,kHz}. 
%In this case, the logarithmic compression is not learnt, as we are restricted to the specifications of the original model~\cite{hershey2017cnn}. 
In order to match the same conditions as previous experiments for inputs longer than 1\,s, we compute several predictions by a non-overlapping moving window of 1\,s that are finally averaged. When audios are shorter than 1\,s, we use `repeat-pad' as in previous experiments (see section~\ref{sec:traindetail}).

In order to study how the pre-trained Audioset model transfers to our tasks, we consider three alternatives:
\begin{itemize}
\item \textbf{Nearest-neighbor with Audioset features:} This baseline classifier serves to study how discriminative are the Audioset features alone. It is based on the cosine distance, and utilizes majority voting to aggregate the different predictions of the model through time (one per second of audio).

\item \textbf{Transfer learning (fine-tuning):} The pre-trained Audioset model is fine-tuned, together with a dense softmax layer that acts as the final classifier. %This can be considered the traditional form of transfer learning. 

\item \textbf{Prototypical networks + transfer learning:} We experiment with the idea of using transfer learning in the context of prototypical networks, and we fine-tune the pre-trained model together with a dense linear layer (10 units) that defines the embedding space where the distance-based classifier operates.
\end{itemize}

For all transfer learning experiments, in order to avoid pre-trained layers to quickly overfit the train set, fine-tuning occurs at a slower pace than training the classification or embedding layer. We use a learning rate of 0.00001 for the pre-trained layers, and a learning rate of 0.1 for randomly initialized layers.

%%%%%%%%%%%%%%%%%%%%%%%%%%%%%%%%%%%%%%%%%%%%%%%%%%%%%%%%%%%%%%%%%%%%%%%%%%%%%%%%%%%%%

\section{Results}
\label{sec:results}

\begin{figure*}[t]
	\centering
	\rotatebox[origin=l]{-90}{\includegraphics[width=0.53\linewidth]{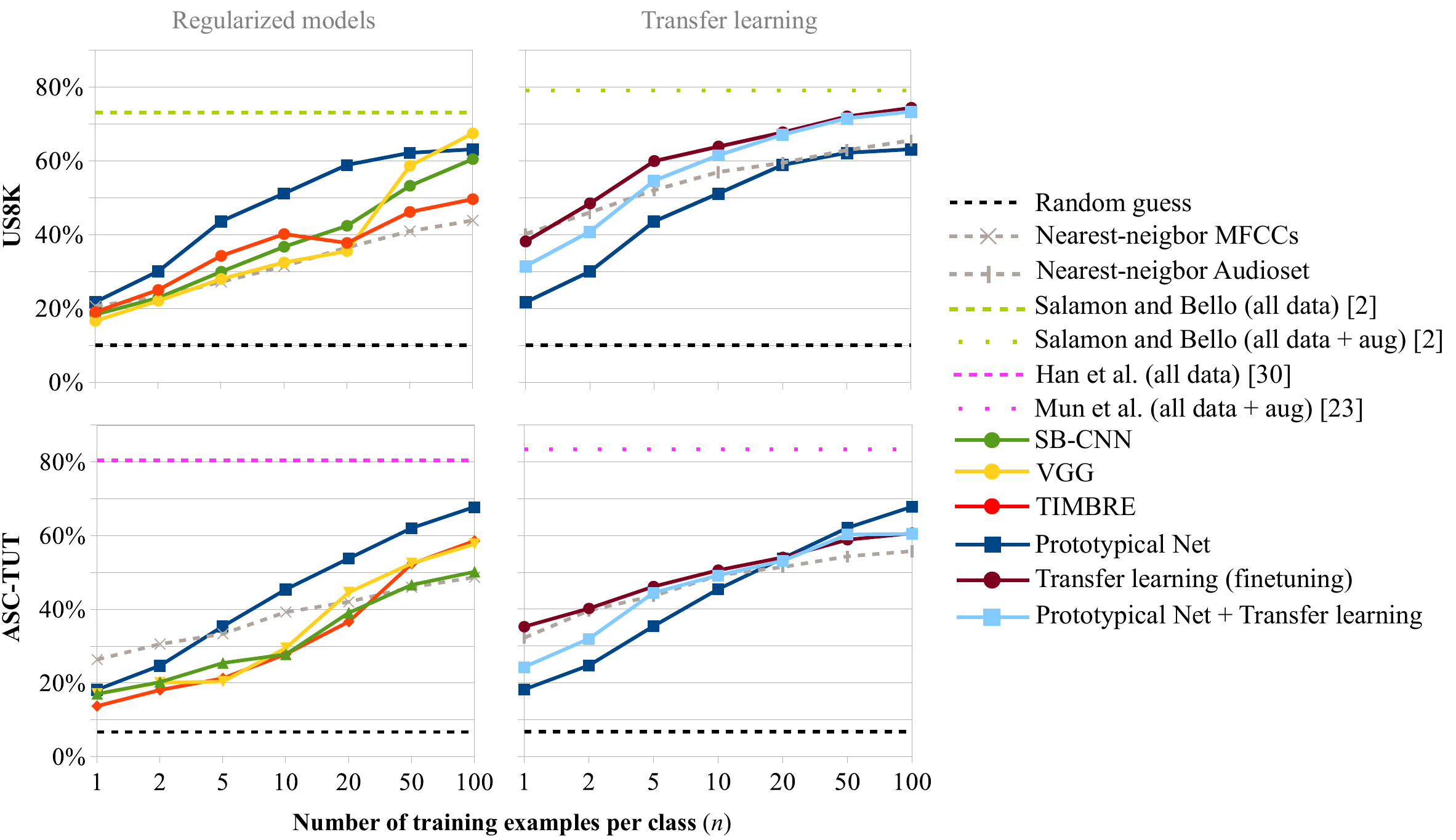}}
    \vspace{-1mm}
	\caption{Accuracy (\%) of the studied strategies when compared to prototypical networks. Dashed and dotted lines represent baselines, and strong lines represent the considered strategies. For comparison, we repeat the curve for prototypical networks in both left and right plots.}
	\label{fig:results}
\end{figure*}

Fig.~\ref{fig:results} summarizes the results we obtain for the two datasets considered in this work: US8K (first row) and ASC-TUT (second row).
On the left-hand side, we compare the results of the regularized models with the ones of prototypical networks. On the right-hand side, we compare the results of transfer learning with the ones of prototypical networks. For each study case, we depict a lower bound (random guess) and an upper bound (previous works using all the train set, see section~\ref{sec:baselines}). In addition, we include a basic baseline, consisting of a nearest neighbor classifier. All baselines and references are depicted with dashed and dotted lines.

First of all, we elaborate on the results obtained by the standard and regularized deep learning models, namely SB-CNN, VGG, and TIMBRE (Fig.~\ref{fig:results}, left). 
%The strongly regularized TIMBRE model, with a single layer CNN having just as many vertical filters as classes, %and  an order of magnitude less number of parameters, 
%achieves equivalent (or better) results than VGG when few training data are available. 
All these models perform similarly when few training data are available ($n<50$). Even so, it is remarkable the performance of the strongly-regularized TIMBRE model for the US8K dataset, as this outperforms the other two and the MFCC baseline when $n\leq 10$.
Notice, however, that the trend changes when more data becomes avilable ($n>20$). Under this data regime, the expressive VGG model seems to better exploit the available data.
%outperforms TIMBRE when more than 50 training examples are available per class. 
Finally, it is also interesting to observe the limitations of the commonly used deep learning pipeline when few training data are available, as the SB-CNN and regularized models struggle to clearly outperform a simple nearest-neighbor MFCC baseline for~$n\leq 20$.

In order to overcome the abovementioned limitations of standard and regularized deep learning models, we investigate the use of prototypical networks. 
Fig.~\ref{fig:results}~(left) depicts how they consistently outperform raw deep learning models, both for US8K and ASC-TUT. Although performance gains are moderate for $n<5$, prototypical networks clearly outperform regularized models for $5< n < 50$ (raw results are available in Appendix~\ref{app:D}). Interestingly, though, the performance of prototypical networks can saturate for $n>50$, as for US8K results. This suggests that prototypical networks may not be as competitive as regular deep learning architectures when sizable amounts of training data are available. However, such tendency could be data-dependent, as it is not observed for ASC-TUT.
%This suggests that, in practice, collecting around 50 examples per class might suffice for building a reasonable audio classifier.

Note that the prototypical networks' embedding space is defined by a larger VGG than the regularized VGG model. Hence, prototypical networks could seem to be more prone to overfitting than regularized models. However, we find that prototypical networks do generalize better than standard and regularized models in the low data regime (Fig.~\ref{fig:results}, left). Since overfitting seems not to dramatically affect prototypical networks' results, we find that highly expressive models, like a large VGG, can deliver good results. We speculate that this might be caused because the resulting latent space is competent enough to discriminate each of the classes, whereas for smaller and less expressive models this might not be the case.

In our study, we also investigate how transfer learning approaches compare with the previous solutions (Fig.~\ref{fig:results}, right). We observe that, %particularly for $n<10$, transfer learning methods clearly outperform the rest. 
for $n\geq10$, transfer learning with basic fine-tuning performs equivalently to prototypical networks + transfer learning. However, for $n<10$, the former outperforms the latter.
We speculate that this effect emerges when prototypes trained with small data are not representative enough of their corresponding classes. 
%For such small data, $n<10$, 
Remember that the support set to compute each class' prototype is conformed by 5 randomly selected patches from the train set. As a result, for example, when $n=1$ these 5 patches are sampled from the same spectrogram and then reused for training~$f_\phi$. Consequently, the variety of the training examples used for computing the prototypes can be very limited for $n<10$, what might be harming the results of prototypical networks. Interestingly, though, for $n\geq 10$ we observe that prototypical networks + transfer learning start performing equivalently to transfer learning with fine-tuning. Note that $n=10$ is the first scenario where prototypical networks can be trained with data being variate enough, since 5 examples can be used for computing the prototypes and 5 additional examples can be used for training $f_\phi$ (see section~\ref{sec:proto}).

Finally, it is worth reminding that source and target tasks are the same for US8K but are different for ASC-TUT. Possibly for that reason, transfer learning consistently outperforms prototypical networks for US8K, but struggles to do so for ASC-TUT. For the latter dataset, we see that prototypical networks (trained from scratch with $n\leq 100$ instances) are able to outperform transfer learning-based approaches (pretrained with 2\,M audios) for $n>20$.
This result denotes that transfer learning has a strong potential when small data are available ($n\leq20$). However, it can be easily overthrown by prototypical networks if the number of training examples per class becomes large enough, and target and source tasks do not match.

\section{Discussion}
\label{sec:discuss}

Among the strategies we have studied for training neural network classifiers with few annotated audios, we have found prototypical networks and transfer learning to be the ones providing the best results. %However, each of these approaches might respond well under different conditions.
However, choosing one or another might depend on the specificities of the use case.
Transfer learning-based classifiers are generally a good choice
%tends to deliver good performance 
when operating in low-data regimes, but they assume that a pre-trained model is readily available. Importantly, such model needs to be trained with data falling under a similar distribution to the few data samples we have available for solving the task. Otherwise, there is no guarantee for transfer learning to deliver better results than, for instance, prototypical networks. 
When data distributions do not match, %or cannot be properly assessed, in our experiments 
we show that prototypical networks trained from scratch can be the right choice. 

In order to restrict ourselves to a realistic low-data scenario, our results are computed without utilizing any validation set. As a result, the set of intuitions that generally help us deciding when to stop training no longer hold. We have found early stopping to be a valid approach when training regular deep learning classifiers. However, interestingly, we have found overfitting not to dramatically affect prototypical networks' results, possibly because these rely on a robust distance-based classifier. Since deciding in which epoch to `early-stop' highly depends on many design choices, we have found the `just overfit' criteria of prototypical networks to be very simple while delivering competitive results.

{%\small
\bibliography{bibliography} 
\bibliographystyle{IEEEbib}
}

%%%%%%%%%%%%%%%%%%%%%%%%%%%%%%%%%%%%%%%%%%%%%%%%%%%%%%%%%%%%%%%%%%%%%%%%%%%%%%%%%%%%%

\clearpage
\onecolumn

\appendix

\section{Appendix}

\subsection{Prototypical networks distance: Euclidean vs.\ cosine}
\label{app:B}
\begin{figure}[!h]
	\centering
    \vspace{-3mm}
	\rotatebox[origin=l]{-90}{\includegraphics[width=0.36\linewidth]{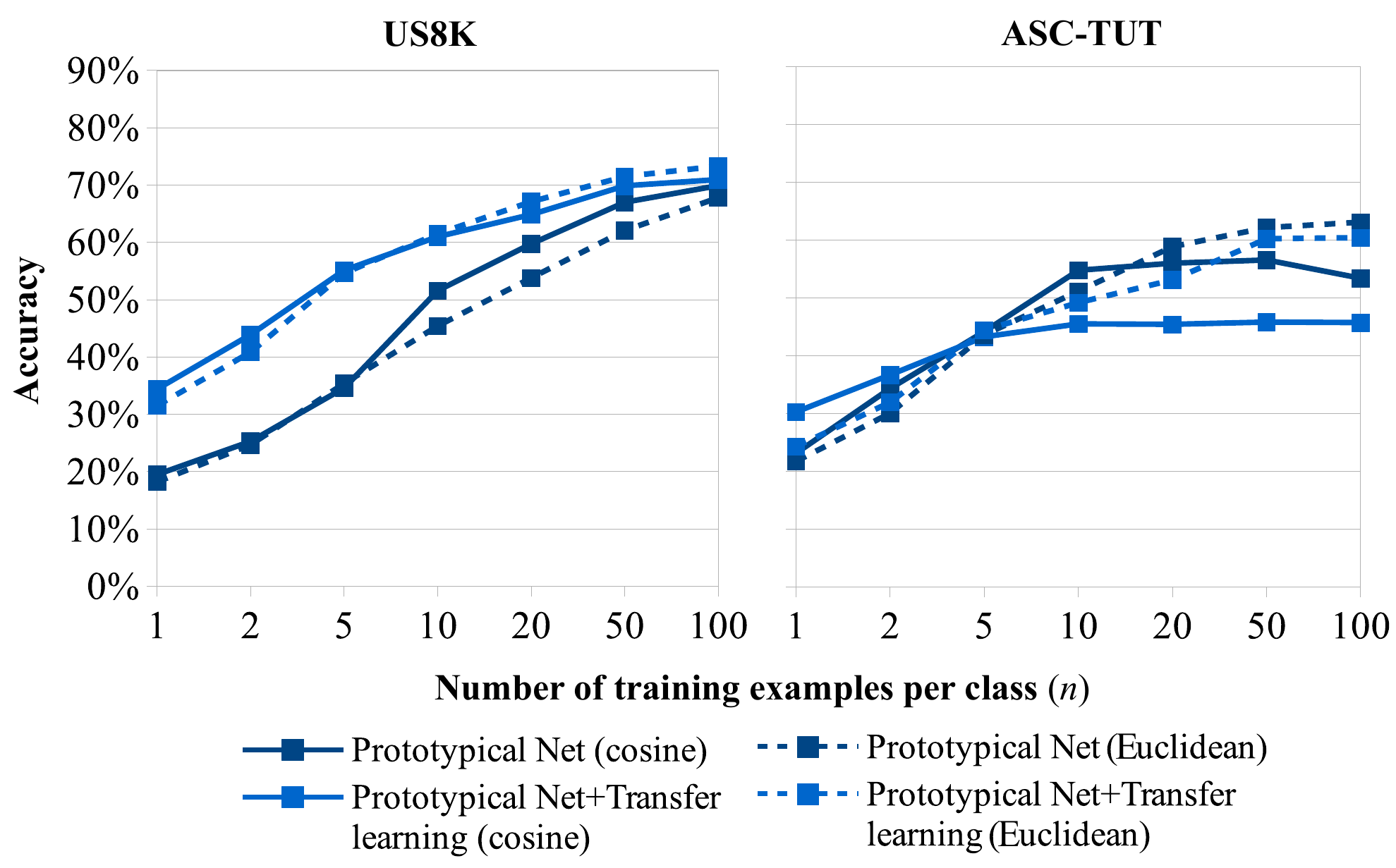}}
	\vspace{-2mm}
	\caption{Accuracy (\%) results comparing prototypical network-based models when using Euclidean or cosine distance.}
	\label{fig:distances}
\end{figure}

While previous researchers \cite{ravi2016optimization,vinyals2016matching} employed the cosine distance for few-shot learning, the original authors of prototypical networks found the Euclidean distance to improve their results~\cite{snell2017prototypical}. In the following, we study the impact of choosing one distance or another for the two datasets considered in our work. We report the accuracy curves for protypical networks (trained from scratch) and prototypical networks + transfer learning when trained with different amounts of data.
Fig.~\ref{fig:distances} (left) depicts US8K results, where we observe that Euclidean- and cosine-based models perform equivalently. However, for $n\geq 20$, cosine-based prototypical networks consistently achieve around 5\% more accuracy.
Fig.~\ref{fig:distances} (right) depicts ASC-TUT results, where Euclidean- and cosine-based models perform similarly for $n\leq10$. However, for $n>10$, Euclidean-based models outperform cosine-based ones for a large margin.  
Consequently, for our experiments, we decide to utilize Euclidean-based prototypical networks, like the original authors of prototypical networks~\cite{snell2017prototypical}.

\subsection{Prototypical networks: Overfitting or generalization?}
\label{app:C}

\begin{figure}[!h]
\centering
    \vspace{-4mm}
\rotatebox[origin=l]{-90}{\includegraphics[width=0.45\linewidth]{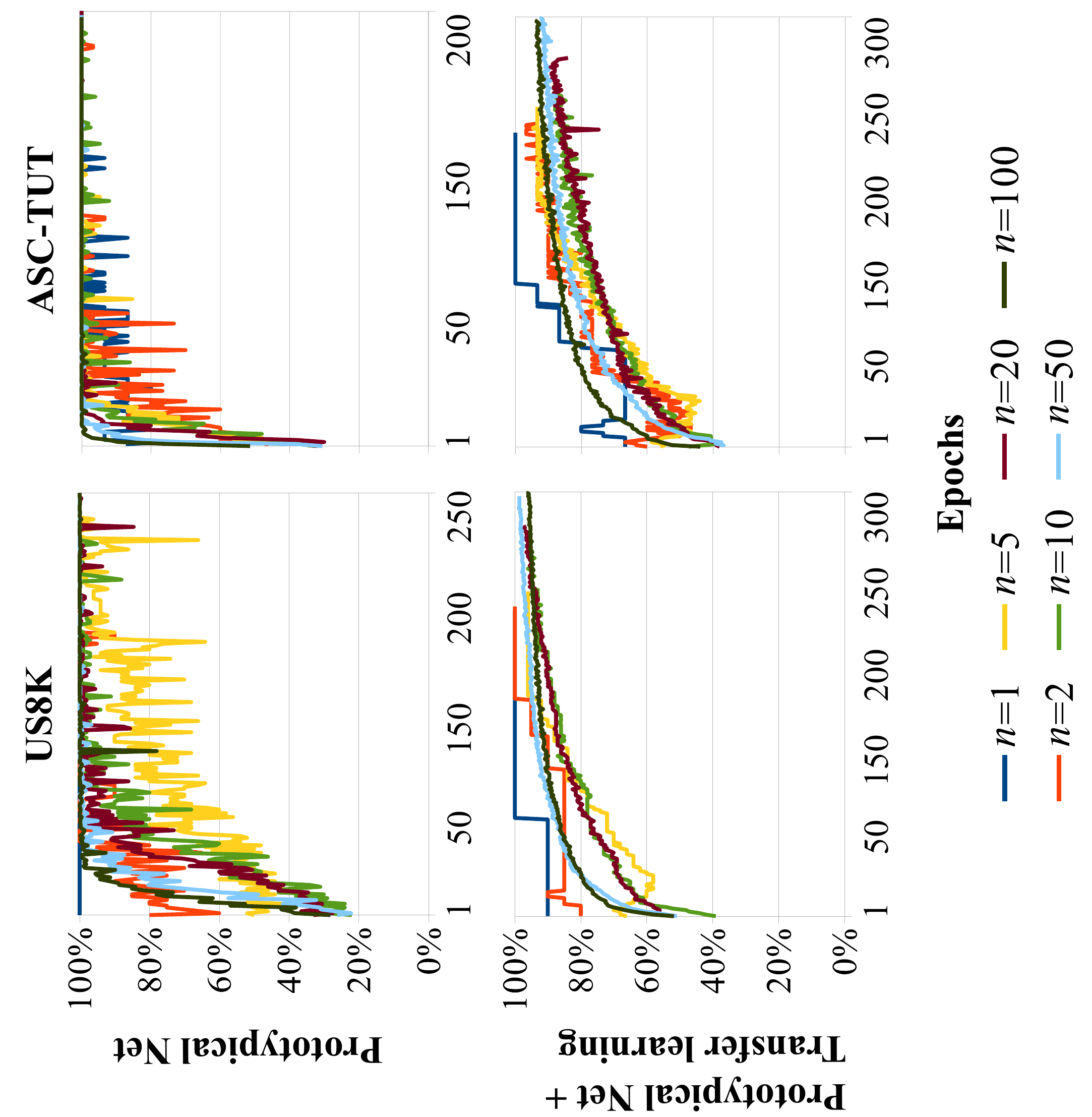}}
\vspace{-3mm}
\caption{Train set accuracy results (\%) for prototypical networks-based models.}
\label{fig:overfit}
\end{figure}

One particularly interesting outcome of our work is that prototypical networks can generalize although they explicitly overfit the train set. In this section, we aim to provide further evidence of this behavior. To this end, we plot the evolution of the train set accuracy during training (we measure train set accuracy in every epoch). For each dataset and prototypical networks-based model, we depict train set accuracy curves when training with different amounts of data. These curves (Fig.~\ref{fig:overfit}) correspond to a single run and were randomly selected.

Although the test set results (depicted in  Fig.~\ref{fig:results}) clearly show that prototypical networks can generalize, it is also manifest from the train set results (depicted in Fig.\ref{fig:overfit}) that prototypical networks do overfit the train set. This effect is particularly notorious when prototypical networks are trained from scratch (no transfer learning), as these tend to quickly overfit. 
In addition, note that the models trained with less data ($n=1$ or $n=2$) tend to overfit quicker, as expected. This fact is particularly noticeable for prototypical networks + transfer learning models, since the used initialization prevents them to quickly overfit. Consequently, the training curves for $n=1$ and $n=2$ are more visible. 

\begin{figure}[!h]
\centering
    \vspace{-2mm}
\rotatebox[origin=l]{-90}{\includegraphics[width=0.45\linewidth]{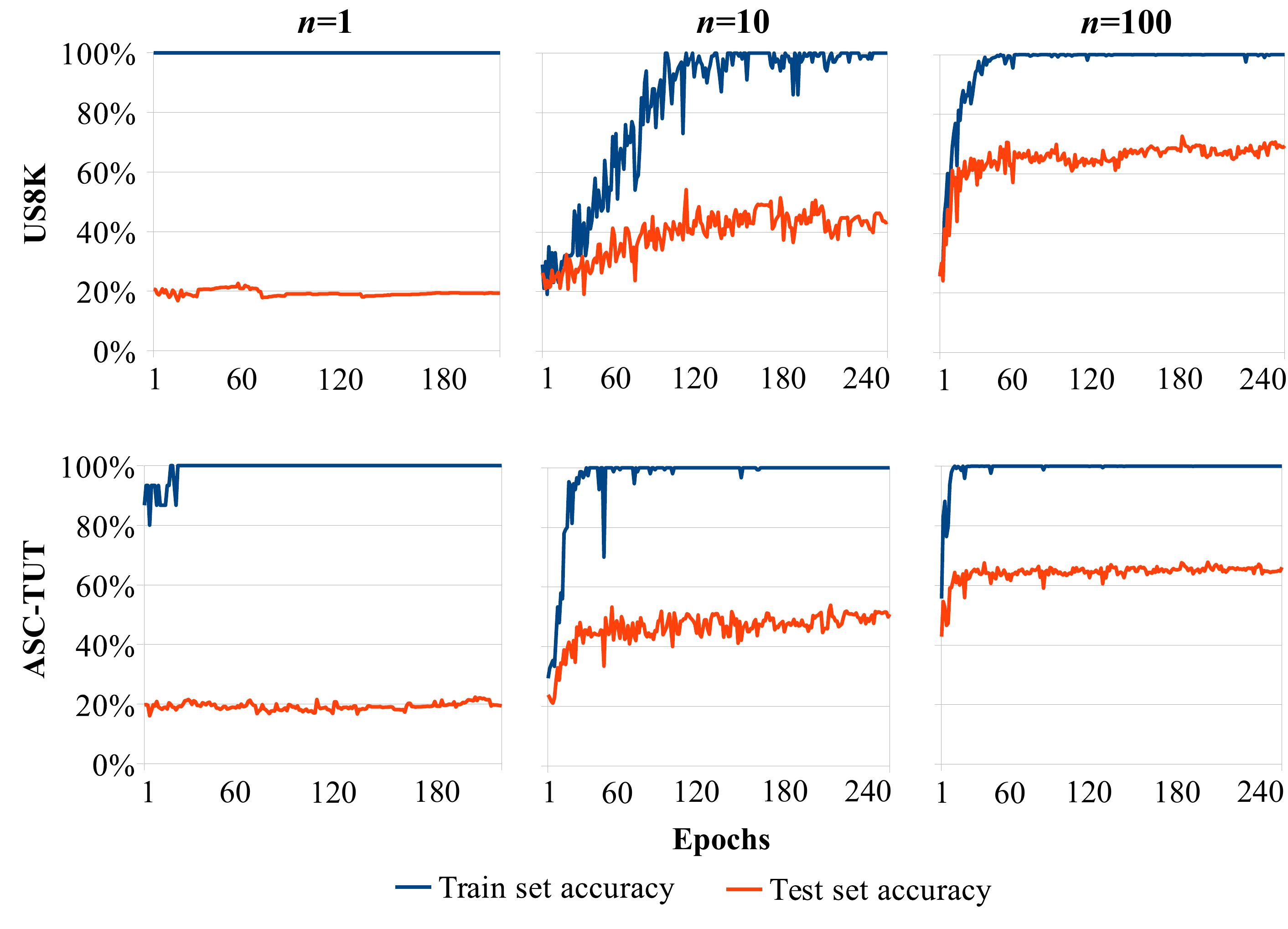}}
\vspace{-3mm}
\caption{Prototypical networks' train set and test set accuracy results (\%) when trained with different amounts of data (different \textit{n}'s).}
\label{fig:overfit_tt}
\end{figure}

In order to further exemplify this behavior, Fig.~\ref{fig:overfit_tt} depicts train set and test set accuracy curves when training for US8K and ASC-TUT with different amounts of data. Train set accuracy curves clearly show how prototypical networks are overfitting the train set (they achieve 100\% train set accuracy). However, interestingly, test set accuracy results do not decrease once overfitting starts. Although one would expect its performance to decrease once overfitting occurs,  we see that prototypical networks' performance remains unaltered.

\subsubsection*{If no validation set is available, when to stop training?} 

The here described `overfitting effect' of prototypical networks is particularly useful when only few data are available for training neural audio classifiers.
As a result of such data constraints, sometimes it can be difficult to assume that a validation set is readily available. Consequently, it might be hard to know when to stop training. However, as seen, prototypical networks can overfit the train set and still deliver sounding results. For that reason, we propose to use the train set accuracy (measured  every epoch) as a proxy to monitor how discriminative is the embedding space, to stop training after the model does not improve its train set accuracy. Note, then, that a discriminative embedding capable to separate all training examples is defined by a model explicitly overfitting the train set. Which, accordingly, would achieve 100\% train set accuracy (a behavior that we repeatedly observe in Fig.~\ref{fig:overfit} and Fig.~\ref{fig:overfit_tt}). Although the stop condition we utilize is encouraging the model to explicitly overfit the train set, prototypical networks can outperform the rest of the models on the test set (see Fig.\ref{fig:results}). Such robustness against overfitting makes prototypical networks particularly convenient for use cases where no validation data is accessible. 

In our work, we decided to stop training after the train set accuracy did not improve for 200 epochs. It is set to 200 for consistency with the rest of the models, since these were trained for 200 epochs (see Section~\ref{sec:sl}). However, note that for prototypical networks this hyper-parameter is much more robust, as for any value greater than 50 results will remain equivalent (see test set accuracy results in Fig.\ref{fig:overfit_tt}).

\subsection{Negative result: Learning the logarithmic compression of the mel spectrogram}
\label{app:A}

Currently, successful neural audio classifiers use log-mel spectrograms as input~\cite{hershey2017cnn,salamon2017deep,pons2017timbre,pons2017end}. Given a mel-spectrogram matrix $X$, the logarithmic compression is computed as follows: $f(x) = \log(\alpha \cdot X + \beta).$
Common pairs of $(\alpha, \beta)$ are $(1, \epsilon)$~\cite{salamon2017deep}, or $(10000, 1)$~\cite{pons2017end}.
In this section, we investigate the possibility of learning $(\alpha, \beta)$. To this end, we investigate two log-mel spectrogram variants:

\begin{itemize}
\item \textbf{Log-learn:} The logarithmic compression of the mel spectrogram $X$ is optimized via SGD together with the rest of the parameters of the model. We use exponential and softplus gates to control the pace of $\alpha$ and $\beta$, respectively. We set the initial pre-gate values to 7 and 1, what results in out-of-gate $\alpha$ and $\beta$ initial values of 1096.63 and 1.31, respectively. 
\item \textbf{Log-EPS:} We set as baseline a log-mel spectrogram which does not learn the logarithmic compression. $(\alpha, \beta)$ are set to $(1, \epsilon)$, as done by Salamon and Bello~\cite{salamon2017deep} for their SB-CNN model.
\end{itemize}

Tables~\ref{table:logus8k} and~\ref{table:logasc} compare the results obtained by several models when varying the mel spectrogram compression: \mbox{log-learn vs. log-EPS}. To clearly illustrate which are the performance gains obtained by log-learn, Tables~\ref{table:logus8k_diff} and \ref{table:logasc_diff} list the accuracy differences between log-learn and log-EPS variants. %caused by learning the logarithmic compression of the mel spectrogram.

\begin{table*}[!h]
	\centering
        %\vspace{-10mm}
	\begin{tabular}{c||cc|cc|cc|cc}
		& \textit{Prototypical} & \textit{Prototypical} & \textit{SB-CNN} & \textit{SB-CNN} & \textit{VGG} & \textit{VGG} & \textit{TIMBRE} & \textit{TIMBRE} \\
		{\textit{n}} & \textit{Net}  (log-learn) & \textit{Net} (log-EPS) & (log-learn) & (log-EPS) & (log-learn) & (log-EPS) & (log-learn) & (log-EPS) \\ \hline
		1 & {19.16}\% & \textbf{21.69}\% & \textbf{19.84}\% & 18.29\% & \textbf{18.27}\% & 16.58\% & \textbf{21.27}\% & 18.97\% \\
		2 & {25.85}\% & \textbf{30.02}\% & \textbf{24.77}\% & 22.81\% & 21.53\% & \textbf{22.03}\% & \textbf{25.95}\% & 24.95\% \\
		5 & {35.85}\% & \textbf{43.58}\% & \textbf{32.85}\% & 29.89\% & 26.48\% & \textbf{27.93}\% & 34.02\% & \textbf{34.20}\% \\
		10 & 44.90\% & \textbf{51.14}\% & \textbf{39.75}\% & 36.66\% & 30.89\% & \textbf{32.40}\% & 39.20\% & \textbf{40.12}\% \\
		20 & 51.15\% & \textbf{58.86}\% & \textbf{45.03}\% & 42.34\% & 32.04\% & \textbf{35.49}\% & \textbf{42.38}\% & 37.70\% \\
		50 & 60.39\% & \textbf{62.14}\% & \textbf{54.94}\% & 53.19\% & 55.22\% & \textbf{58.62}\% & \textbf{48.40}\% & 46.11\% \\
		100 & \textbf{65.38}\% &{63.08}\% & 60.34\% & \textbf{60.43}\% & 64.65\% & \textbf{67.41}\% & \textbf{50.14}\% & 49.57\% \\
	\end{tabular}
	\caption{US8K dataset: accuracy results comparing log-EPS (standard log-mel spectrogram) \& log-learn (learned log-mel spectrogram).}
		\label{table:logus8k}
\end{table*}

\begin{table}[!h]
	\centering
	\begin{tabular}{c||c|c|c|c}
		& \textit{Prototypical}  &  &  & \\
		{\textit{n}} & \textit{Networks} &\textit{SB-CNN} &  \textit{VGG} & \textit{TIMBRE} \\ \hline
		1& \textbf{-2.53}\% & 1.55\% & 1.69\% & 2.30\%\\
		2& \textbf{-4.17}\% & 1.96\% & \textbf{-0.50}\% & 1.00\% \\
		5& \textbf{-7.73}\%& 2.96\% & \textbf{-1.45}\% & \textbf{-0.18}\% \\
		10& \textbf{-6.24}\% & 3.09\% & \textbf{-1.51}\% & \textbf{-0.92}\% \\
		20& \textbf{-7.71}\%& 2.69\% & \textbf{-3.45}\% & 4.68\%\\
		50& \textbf{-1.75}\%& 1.75\% & \textbf{-3.40}\% & 2.29\% \\
		100& 2.30\%& \textbf{-0.09}\% & \textbf{-2.76}\% & 0.57\% \\
	\end{tabular}
	\caption{US8K dataset: log-learn accuracy gains when compared to log-EPS.}
	\label{table:logus8k_diff}
\end{table}

Tables~\ref{table:logus8k} and~\ref{table:logus8k_diff} reveal that log-lean and log-EPS results are almost identical for US8K. Although it seems that log-learn can help improving the results for SB-CNN and TIMBRE architectures, for prototypical networks and VGG one can achieve worse results. For this reason, we conclude that log-learn and log-EPS results are almost equivalent for US8K.
However, for ASC-TUT dataset, log-learn results are much worse than log-EPS ones. Tables~\ref{table:logasc} and~\ref{table:logasc_diff} show that log-learn only improves the results of SB-CNN models when trained with little data ($1\leq n\leq10$), but for the rest of the models the performance decreases substantially.
Accordingly, we decide not to learn the logarithmic compression of the mel spectrogram throughout our study.

\begin{table*}[!h]
	\centering
	\begin{tabular}{c||cc|cc|cc|cc}
		& \textit{Prototypical} & \textit{Prototypical} & \textit{SB-CNN} & \textit{SB-CNN} & \textit{VGG} & \textit{VGG} & \textit{TIMBRE} & \textit{TIMBRE}  \\
		{\textit{n}} & \textit{Net} (log-learn) & \textit{Net} (log-EPS) & (log-learn) & (log-EPS) & (log-learn) & (log-EPS) & (log-learn) & (log-EPS) \\ \hline
		1 & \textbf{19.94}\% & {18.16}\% & \textbf{18.69}\% & 13.70\% & 11.07\% & \textbf{17.01}\% & 16.06\% & \textbf{17.00}\% \\
		2 & \textbf{24.72}\% & {24.68}\% & \textbf{21.34}\% & 18.08\% & 11.54\% & \textbf{20.05}\% & 19.72\% & \textbf{20.21}\% \\
		5 & 31.77\% & \textbf{35.36}\% & \textbf{26.62}\% & 21.24\% & 12.47\% & \textbf{20.36}\% & 21.88\% & \textbf{25.40}\% \\
		10 & 39.64\% & \textbf{45.39}\% & \textbf{30.22}\% & 27.81\% & 13.71\% & \textbf{29.45}\% & 24.17\% & \textbf{27.74}\% \\	 
		20 & 43.28\% & \textbf{53.78}\% & 34.87\% & \textbf{36.61}\% & 22.51\% & \textbf{44.58}\% & 25.09\% & \textbf{39.02}\% \\
		50 & 45.43\% & \textbf{62.03}\% & 45.27\% & \textbf{52.32}\% & 37.14\% & \textbf{52.46}\% & 31.66\% & \textbf{46.61}\% \\
		100 & 47.70\% & \textbf{67.78}\% & 52.59\% & \textbf{58.56}\% & 42.53\% & \textbf{57.71}\% & 38.35\% & \textbf{50.16}\% \\
	\end{tabular}
	\caption{ASC-TUT dataset: accuracy results comparing log-EPS (standard log-mel spectrogram) \& log-learn (learned log-mel spectrogram).}
    \label{table:logasc}
\end{table*}

\begin{table}[!h]
	\centering
	\begin{tabular}{c||c|c|c|c}
		& \textit{Prototypical}  &  &  & \\
		{\textit{n}} & \textit{Networks} &\textit{SB-CNN} &  \textit{VGG} & \textit{TIMBRE} \\ \hline
		1& 1.78 & 4.99 & \textbf{-5.94} & \textbf{-0.94}\\
		2& 0.04 & 3.26 & \textbf{-8.51} & \textbf{-0.49}\\
		5& \textbf{-3.59} & 5.38 & \textbf{-7.89} & \textbf{-3.52}\\
		10& \textbf{-5.70} & 2.41& \textbf{-15.74} & \textbf{-13.93} \\
		20& \textbf{-10.50} & \textbf{-1.74} & \textbf{-22.07} & \textbf{-13.93} \\
		50& \textbf{-16.60} & \textbf{-7.05} & \textbf{-15.32} & \textbf{-14.95} \\
		100& \textbf{-20.08} & \textbf{-5.97} & \textbf{-15.18} & \textbf{-11.81} \\
	\end{tabular}
	\caption{ASC-TUT dataset: log-learn accuracy gains when compared to log-EPS.}
	\label{table:logasc_diff}
\end{table}

\newpage

\subsection{Raw accuracy results}
\label{app:D}

Tables 5 and 6 list the raw accuracy results of the main figure of the paper (Fig.~\ref{fig:results}).

\begin{table*}[ht!]
	\centering
	\begin{tabular}{c||c|cccc||c|cc}
		&  & &  &  & \textbf{Prototypical} &  & \textbf{Transfer learning}  & \textbf{Prototypical Net+}  \\
		\textit{n} & \textbf{NN-MFCC} & \textbf{SB-CNN} & \textbf{VGG} & \textbf{TIMBRE} & \textbf{Net} & \textbf{NN-Audioset} & \textbf{(fine-tuning)} & \textbf{Transfer learning}  \\ \hline
		1 & 20.57\% & 18.30\% & 16.58\% & 18.98\% & 21.69\% & 40.17\% & 38.15\% & 31.48\%  \\
		2 & 23.03\% & 22.81\% & 22.03\% & 24.95\% & 30.02\% & 46.00\% & 48.44\% & 40.82\%  \\
		5 & 27.15\% & 29.89\% & 27.94\% & 34.21\% & 43.58\% & 52.02\% & 59.89\% & 54.61\%  \\
		10 & 31.40\% & 36.66\% & 32.41\% & 40.12\% & 51.14\% & 56.91\% & 63.81\% & 61.48\%  \\
		20 & 36.45\% & 42.34\% & 35.49\% & 37.70\% & 58.86\% & 59.42\% & 67.64\% & 67.07\%  \\
		50 & 40.89\% & 53.19\% & 58.62\% & 46.11\% & 62.14\% & 62.85\% & 71.95\% & 71.47\%  \\
		100 & 43.81\% & 60.43\% & 67.41\% & 49.57\% & 63.08\% & 65.47\% & 74.26\% & 73.28\%  \\
	\end{tabular}
	
	\caption{US8K dataset: Fig.\ref{fig:results} (top) raw results.}
\end{table*}

\begin{table*}[ht!]
	\centering
	\begin{tabular}{c||c|cccc||c|cc}
		&  & &  &  & \textbf{Prototypical} &  & \textbf{Transfer learning}  & \textbf{Prototypical Net+}  \\
	    \textit{n} & \textbf{NN-MFCC} & \textbf{SB-CNN} & \textbf{VGG} & \textbf{TIMBRE} & \textbf{Net} & \textbf{NN-Audioset} & \textbf{(fine-tuning)} & \textbf{Transfer learning}  \\ \hline
		1 & 26.33\% & 13.70\% & 17.01\% & 17.00\% & 18.16\% & 32.18\% & 35.18\% & 24.25\%\\
		2 & 30.52\% & 18.08\% & 20.05\% & 20.21\% & 24.68\% & 39.58\% & 40.13\% & 31.91\%\\
		5 & 33.31\% & 21.24\% & 20.36\% & 25.40\% & 35.36\% & 43.56\% & 46.09\% & 44.41\%\\
		10 & 39.21\% & 27.81\% & 29.45\% & 27.74\% & 45.39\% & 49.09\% & 50.53\% & 49.15\%\\
		20 & 41.99\% & 36.61\% & 44.58\% & 39.02\% & 53.78\% & 51.39\% & 54.00\% & 53.09\%\\
		50 & 45.98\% & 52.32\% & 52.46\% & 46.61\% & 62.03\% & 54.28\% & 58.79\% & 60.22\%\\
		100 & 48.66\% & 58.56\% & 57.71\% & 50.16\% & 67.78\% & 55.69\% & 60.56\% & 60.39\%\\
	\end{tabular}
	
	\caption{ASC-TUT dataset: Fig.\ref{fig:results} (bottom) raw results.}
\end{table*}

\end{document}